\documentclass[12pt]{article}
\usepackage{graphicx,multicol}
\usepackage{amssymb,amsmath}
\usepackage[usenames]{color}
\usepackage{bbold}
\DeclareMathOperator{\Tr}{Tr}                % trace operator
\DeclareMathOperator{\Pf}{Pf}

\newcommand{\Ncal}{\mathcal{N}}

\newcommand{\Ocal}{\mathcal{O}}

\usepackage[normalem]{ulem}

\def\be{\begin{equation}}
\def\ee{\end{equation}}  

\begin{document}
\begin{titlepage}
\hfill
\vbox{
    \halign{#\hfil         \cr
           MCTP-10-13  \cr
           } % end of \halign
      }  % end of \vbox
\vspace*{20mm}
\begin{center}
{\Large \bf First Results from Lattice Simulation of the PWMM}

\vspace*{15mm}
\vspace*{1mm}

{Simon Catterall$^1$ and Greg van Anders$^2$}

\vspace*{1cm}

{$^1$ Department of Physics, Syracuse University, Syracuse, NY 13244, USA\\
$^2$ Michigan Center for Theoretical Physics, Randall Laboratory of Physics,\\
The University of Michigan, Ann Arbor, MI 48109-1040, USA}

\vspace*{1cm}
%%\maketitle
\end{center}

\begin{abstract}
  We present results of lattice simulations of the Plane Wave Matrix Model
  (PWMM). The PWMM is a theory of supersymmetric quantum mechanics that has a
  well-defined canonical ensemble. We simulate this theory by applying rational
  hybrid Monte Carlo techniques to a na\"ive lattice action. We examine the
  strong coupling behaviour of the model focussing on the deconfinement
  transition.
\end{abstract}

\end{titlepage}
\section{Introduction} \label{sec:intro}
The AdS/CFT correspondence \cite{adscft1,adscft2,adscft3} has emerged as a
useful setting for studying certain strongly coupled gauge theories. Though
this correspondence has been widely tested, it is useful to look for examples
in which the correspondence can be tested directly by studying theories in
which tractable calculations can be done directly at strong coupling. Lower
dimensional examples would seem to provide a useful setting for such a study.
Recent effort has been devoted to studying the BFSS matrix model at strong
coupling using both lattice and non-lattice techniques \cite{hnt,ahnt,cw2}.
However, even that one dimensional quantum theory is still not well defined in
the canonical ensemble because it has a moduli space \cite{cw3}. The moduli
space in this theory is given by the eigenvalues of nine commuting matrices
that transform under the $SO(9)$ R-symmetry. The existence of these flat
directions means that, at finite temperature, the partition function is formally
divergent, and it was shown in \cite{cw3} that when Monte Carlo simulations of
this theory are performed, this divergence eventually causes the simulation to
break down.

The Plane Wave Matrix Model (PWMM) \cite{bmn} arises as a natural next choice
for simulation because it includes supersymmetry preserving mass terms that lift
the moduli space, and give a discrete set of vacua. The theory still has enough
supersymmetry that it has a known gravity dual at zero temperature \cite{lm}.
Symmetry reduces the dual supergravity problem to a two dimensional one, and
the geometries dual to the discrete set of field theory vacua can be put in
one-to-one correspondence with a set of axisymmetric electrostatics
configurations, the potential of which determines the supergravity fields,
and was solved in \cite{inst}. Moreover, this theory has a number of large
$N$ limits that give theories of physical interest: M-theory on a plane-wave
\cite{bmn}, little string theory on $S^5$ \cite{lmsvv}, and $\Ncal=4$ SYM on
$R\times S^3$ \cite{N4PWMM}. Studying the PWMM at strong coupling, therefore, is
a first step toward direct simulation of these theories.

The PWMM is also a good candidate for establishing benchmarks for studies of
lattice supersymmetry in general. It is simple enough to allow for direct study
using a na\"ive lattice action, is well-defined in the canonical ensemble and of
course being one dimensional is numerically very tractable. As such it is
complementary to studies of supersymmetric theories in higher dimensions using
new lattice formulations which retain some exact supersymmetry at non zero
lattice spacing - see the recent review \cite{Phys_Rep_Exact_SUSY}. For example
it should be possible to cross check the results of direct simulations of
$\Ncal=4$ super Yang-Mills in certain limits with results derived from
simulations of na\"ive lattice discretizations of the PWMM.

In this paper we will present the results of simulations of the PWMM on the
lattice. We use a na\"ive lattice action, and concentrate on studying the
Hagedorn/deconfinement transition in the model. We study the theory at fixed
temperature, measured in units of the mass deformation, as a function of the
't Hooft coupling (measured in the same units). We used a quenched approximation
to explore the dependence of the critical behaviour on the rank of the gauge
group and the number of lattice points. The quenched approximation is much less
computationally demanding and can be used to estimate reasonable choices of
parameters for simulating the full theory. We establish that we can get a
reasonable approximation of the continuum large $N$ behaviour with modest values
of both $N$ and the size of the lattice. We then simulate the full theory. The
Hagedorn/deconfinement transition in the PWMM has also been studied previously
at weak coupling 't Hooft coupling in the large $N$ limit, and we compare our
results to extrapolations of those from weak coupling.

The remainder of this paper is organized as follows. In section \ref{sec:PWMM}
we review the PWMM, its vacuum solutions, and various limits. In section
\ref{sec:lattice} we present our lattice action and discuss the simulation
method. Section \ref{sec:hag} contains our main results. We review the weak
coupling results and contrast them with the results of our simulations. In
section \ref{sec:disc} we discuss possible completions of the phase diagram.

\section{The PWMM}\label{sec:PWMM}
The Plane-Wave Matrix Model (PWMM) is a gauged matrix quantum mechanics theory
with sixteen supercharges. It has an action that can be written as
\begin{equation}\label{action}
\begin{split}
S=\frac{N}{2\lambda}\int_0^\beta dt
\Tr\Bigl(&-\sum_i(DX_i)^2 -\sum_{i<j}[X^i,X^j]^2
+\Psi^T\gamma^0D\Psi + \sum_i\Psi^T\gamma_i[X^i,\Psi]\\&
-\mu^2\sum_{i=1}^3(X^i)^2 -\frac{\mu^2}{4}\sum_{i=4}^9(X^i)^2
-2\sqrt{2}\mu\epsilon_{ijk}X^iX^jX^k +\frac{3\mu}4\Psi^T\gamma_{123}\Psi
\Bigr),
\end{split}
\end{equation}
in which $X^i$ are nine scalar matrices that sit in the adjoint representation
of $SU(N)$, i.e.\ $X^i=\sum_{a=1}^{N^2-1} X_i^a T^a$, where $T^a$ are
anti-Hermitian generators that are normalized to $\Tr T^aT^b=-\delta_{ab}$;
$\Psi$ is a Majorana fermion that is also in the adjoint representation of the
gauge group; $\Tr$ is over the gauge indices; $D$ is a gauge covariant
derivative; $\lambda$ is the 't Hooft coupling, given by
$\lambda = g^2N$; 
and
\begin{equation}
\gamma_{123}=\frac16\epsilon_{ijk}\gamma_i\gamma_j\gamma_k \, .
\end{equation}

The terms in the first line of \eqref{action} are those in the BFSS matrix
model \cite{bfss}. The terms in the second line give supersymmetry preserving
masses to the fields and add a Myers term. The addition of these terms gives
this model two technical advantages compared with BFSS. Firstly, in the limit
$\mu\to\infty$ the model becomes weakly coupled and can be studied
perturbatively \cite{dsv1}. Also, the addition of these terms lifts the moduli
space of the BFSS model giving instead a discrete set of vacua \cite{bmn}. This
is important for numerical simulation because it means the theory is
well-defined in the canonical ensemble. If we consider the scalar fields $X^i$
for $i=1,2,3$ we can write the potential as
\begin{equation}
  V=-\Tr\left[(\mu X^i+\frac1{\sqrt 2}\epsilon_{ijk} [X^j,X^k])^2\right],
\end{equation}
which is minimized when the scalar fields $X^i$ are proportional to generators
of $SU(2)$. The discrete vacua correspond to the various ways of forming
$N\times N$ matrices that generate $SU(2)$, and can be put in one-to-one
correspondence with partitions of $N$ into a sum of natural numbers. It was
shown in \cite{dsv2} that the vacua sit in certain doubly atypical
representations of the $SU(2|4)$ supersymmetry algebra, which protect them from
quantum corrections, and thus persist at strong coupling.

The identification between the field theory vacuum states and regular type IIA
supergravity solutions with $SU(2|4)$ supersymmetry was found in \cite{lm}.
$SU(2|4)$ contains a bosonic $R\times SO(3)\times SO(6)$ that reduces the type
IIA supergravity problem to a two-dimensional one. Imposing supersymmetry and
regularity reduces the two-dimensional problem to solving a two-dimensional
Laplace equation. The vacua of the PWMM are encoded by choosing appropriate
boundary conditions. Using the fact that the potential in electrostatics obeys
the Laplace equation, it was shown \cite{lm} that regular supergravity solutions
correspond to axisymmetric configurations of charged conducting discs. This
class of electrostatics problems was solved in \cite{inst}.

The PWMM has a number of large $N$ limits in which it describes interesting
physics. See \cite{lmsvv} for a summary; we will mention a few here
here.\footnote{For links between the PWMM and other field theories see
\cite{itt,istt}.}
\begin{enumerate}
\item If we consider a limit in which
\begin{equation}\label{Mlim}
  N\to\infty \qquad \frac{g^2}{N^3} \; \text{fixed},
\end{equation}
this model should describe M-theory in a maximally supersymmetric plane-wave
background. This is because the matrix theory conjecture suggests this model
describes the DLCQ of M-theory on the plane wave with compact null momentum
given by
\begin{equation}
  \mu l_p^2 p^+ = \frac{N}{g^\frac23} \propto \frac{N}{R},
\end{equation}
where $R$ is the size of the null circle.
The limit \eqref{Mlim} decompactifies the compact null circle.
\item The 't Hooft limit, in which $N\to\infty$ with $\lambda$ fixed, allows for
  the perturbative study of the theory if $\lambda$ is small. It is in this
  limit that the Hagedorn/deconfinement transition can be studied analytically,
  as we will review below.
\item It was shown in \cite{lmsvv} that if we consider expanding around the
trivial vacuum solution and take a limit in which, asymptotically,
\begin{equation}
  N\to\infty \qquad \frac{\ln^4 N}{\lambda} \; \text{fixed},
\end{equation}
the PWMM describes type IIA little string theory on $S^5$. Little string theory
is the theory that describes the degrees of freedom on NS5-branes, and is of
particular interest because it exhibits some of the features of string theories
(e.g.\ T-duality), but not others (e.g.\ gravity).
\item Using arguments from the gravity side \cite{lm} or from the field theory
  side \cite{N4PWMM} $\Ncal=4$ SYM on $R\times S^3$ can be realized by expanding
  about a particular vacuum of the PWMM. (For a full discussion, see also
  \cite{N4PWMM2}, \cite{N4PWMM3}.) Expanding about a vacuum consisting of
  representations of $SU(2)$ of dimension $n+j$, for some integers $n$, and
  $j=1,2,3,\dotsc,m$, each with $k$ copies. If the PWMM has gauge group
  $SU(N)$, then we must have
  \begin{equation}
  N=\frac12 km(2n+m+1) .
  \end{equation}
  Taking a limit $N\to\infty$ in which
  \begin{equation}
    n\to\infty \qquad m\to\infty \qquad \frac mn\to0 ,
  \end{equation}
  gives $\Ncal=4$ SYM on $R\times S^3$ with gauge group $SU(k)$.
\end{enumerate}

It is in principle possible to extract the physics of these models by studying
the behaviour of the model along appropriate curves in the $\lambda-N$ plane,
and extrapolating to large $N$. In this paper we seek to establish benchmarks
for this procedure by performing a lattice simulation of the PWMM at strong
coupling and finite $N$.

\section{Lattice Action}\label{sec:lattice}
To find our lattice action, we start with the PWMM action \eqref{action}.
The bare action has three parameters with unit mass dimension 
$\lambda^{\frac13}$, $\beta^{-1}$ and $\mu$. The gauge field and scalars $X^i$
also have unit mass dimension, and the fermions have mass dimension $\frac32$.
We would like to make the following rescaling:
\begin{equation}
X\to \mu T X \quad D\to \mu T D \quad t \to \frac{t}{\mu T} \quad
\Psi \to (\mu T)^\frac32 \Psi ,
\end{equation}
where $T$ is a dimensionless number that we will shortly take to be the number
of lattice points.  We can then write the action as
\begin{equation}
\begin{split}
S=\frac{\mu^3 T^3 N}{2\lambda}\int_0^{\mu\beta T} dt
\Tr\Bigl(&-\sum_i(DX_i)^2 -\sum_{i<j}[X^i,X^j]^2
+\Psi^T\gamma^0D\Psi + \sum_i\Psi^T\gamma_i[X^i,\Psi]\\&
-\frac1{T^2}\sum_{i=1}^3(X^i)^2 -\frac1{4T^2}\sum_{i=4}^9(X^i)^2
-2\sqrt{2}\epsilon_{ijk}X^iX^jX^k +\frac3{4T}\Psi^T\gamma_{123}\Psi
\Bigr).
\end{split}
\end{equation}
Note that all of our fields have been rendered dimensionless. After integrating
over the fermions we have the following partition function
\begin{equation}
Z=\int dA dX \Pf(\Ocal) e^{-S_B},
\end{equation}
with
\begin{equation}
\begin{split}
S_B=\frac{\mu^3 T^3 N}{2\lambda}\int_0^{\mu\beta T} dt
\Tr\Bigl(&-\sum_i(DX_i)^2 -\sum_{i<j}[X^i,X^j]^2\\&
-\frac1{T^2}\sum_{i=1}^3(X^i)^2 -\frac1{4T^2}\sum_{i=4}^9(X^i)^2
-2\sqrt{2}\epsilon_{ijk}X^iX^jX^k
\Bigr),
\end{split}
\end{equation}
and
\begin{equation}
\Ocal=\gamma^0D+\sum_i\gamma_i[X^i,\cdot]+\frac3{4T}\gamma_{123} .
\end{equation}

We discretize the theory on a lattice with spacing $a$, and replace the
$\int dt$ with $a\sum_t$. We use the derivative operator
\begin{equation}
D = \begin{pmatrix} 0& D^+\\D^- &0\end{pmatrix} ,
\end{equation}
where
\begin{equation}
D^+\Psi = \frac1a(U(t)\Psi(t+a)U^\dagger-\Psi(t)) .
\end{equation}
$D^-$ is the adjoint of $D^+$. The lattice action is
\begin{equation}
\begin{split}
S_B=\frac{\beta\mu^4 T^3 N}{2\lambda}\sum_{m=0}^{T-1}
\Tr\Bigl(&-\sum_i{(DX^i)_m}^2 -\sum_{i<j}[X^i_m,X^j_m]^2\\&
-\frac1{T^2}\sum_{i=1}^3(X^i_m)^2 -\frac1{4T^2}\sum_{i=4}^9(X^i_m)^2
-2\sqrt{2}\epsilon_{ijk}X^i_mX^j_mX^k_m
\Bigr),
\end{split}
\end{equation}
and
\begin{equation}
  \Ocal=\gamma^0D_{mn}+\sum_i\gamma_i[X^i_m,\cdot]\mathbb{1}_{mn}
  +\frac3{4T}\gamma_{123}\mathbb{1}_{mn} .
\end{equation}
The overall dimensionless lattice coupling is then
\begin{equation}
\kappa = \frac{\beta\mu^4T^3N}{2\lambda} ,
\end{equation}
where $\beta$ is the inverse temperature, and $T$ is the number
of lattice points. The continuum limit is defined by taking $T\to\infty$ with
$\frac{\mu^4 N \beta}{2\lambda}$ fixed.

We simulate this model using Rational Hybrid Monte Carlo (RHMC) \cite{rhmc}.
Since the fermions appeared quadratically, we were able to formally integrate
them out yielding the Pfaffian of the operator $\Ocal$. Using the fact that
$\Pf^2(\Ocal)=\det\Ocal$, the Pfaffian can be computed by introducing a path
integral over pseudofermions $F$
\begin{equation}
\Pf\Ocal = \int [DF] e^{-F^\dagger(\Ocal^\dagger\Ocal)^{-\frac14} F}.
\end{equation}
It is still challenging to compute $(\Ocal^\dagger\Ocal)^{-\frac14}$, but it
can be done efficiently by making a rational approximation, and making use of
iterative Krylov subspace methods to solve the resulting set of shifted linear
systems \cite{jegerlehner}. Fictitious conjugate momenta are introduced for each
of the variables, and the resulting system is simulated by using a combination
of molecular dynamics integrations of the equations of motion and Metropolis
Monte Carlo \cite{metropolis,hmc}. The main computational expense derives from
simulating the fermionic Pfaffian, and so it will be useful at times to consider
the quenched approximation (without fermions). This will allow us to estimate
what happens at much larger values of $N$ than we can when the fermions are
included with equivalent computational effort.

\section{Hagedorn/Deconfinement Transition}\label{sec:hag}
To get an idea of how well our finite $N$ simulations work we will compute a
simple observable in the matrix model: the Polyakov loop. There are good reasons
to compute it.
\begin{itemize}
\item It is a simple gauge invariant observable that is straightforward to
define on the lattice.
\item It is an order parameter for deconfinement, and is expected to signify
a geometric transition in the dual gravity theory.
\item It has been computed in the PWMM at weak coupling.
\end{itemize}

In lattice gauge theory, holonomies of the gauge field are natural observables.
In our one-dimensional theory, the holonomy to compute is the one around
the thermal circle. It can serve as an order parameter for deconfinement
because it measures the energy cost of adding an infinitely massive external
charged particle. In the confined phase, the trace of the Polyakov loop has
vanishing expectation value. This is because it is a unitary matrix, and in the
confined phase its eigenvalues are uniformly distributed on the unit circle. In
the deconfined phase this symmetry is broken, the eigenvalues begin to clump and
the Polyakov loop develops a vev.

It is also interesting from the point of view of gauge/gravity duality. In
gravity, there is a well-known phase transition between thermal AdS space and
asymptotically AdS-Schwarzschild black holes \cite{hp}. Black hole
thermodynamics implies that this phase transition comes with an increase in
entropy. It was argued \cite{wittenconf} that this fact, along with the fact
that the Wilson line computed via gauge/gravity duality ceases to obey an area
law in the high temperature phase, means this bulk gravity transition
corresponds to deconfinement in the gauge theory. The increase in gravitational
entropy also matches nicely with this picture because in the confined phase the
free energy is expected to be of order one, whereas in the deconfined phase it
is expected to be of order $N^2$. For a discussion of general aspects of this,
see \cite{ammpv}. We therefore expect that by studying the deconfinement
transition at strong coupling in the matrix model we are capturing a geometric
transition in the bulk dual.

In \cite{ammpv,fss} it was shown that in certain large $N$ gauge theories,
including the PWMM, it is possible to have deconfinement transitions at weak
coupling. Let us discuss briefly why this occurs, using an argument from
\cite{ammpv}. Consider a theory of two bosonic matrix oscillators with unit
energy in the adjoint representation of a gauge group with rank $N$. Gauge
invariant states are formed by the action of traces of matrix creation operators
on the vacuum. Let us consider single trace states, which dominate in the
large $N$ limit. The number of single trace states with energy $E$ is bounded
above by $2^E$, because there are $E$ positions in the trace that can be filled
by either oscillator. Some of these states are related by the cyclicity of the
trace, but there are at most $E-1$ such states, and so the number of states
with energy $E$ is bounded below by $\frac{2^E}E$. This lower bound is
sufficient to produce an exponentially growing density of states, which
indicates that there will be a Hagedorn transition.

\begin{figure}
  \begin{center}
  \input{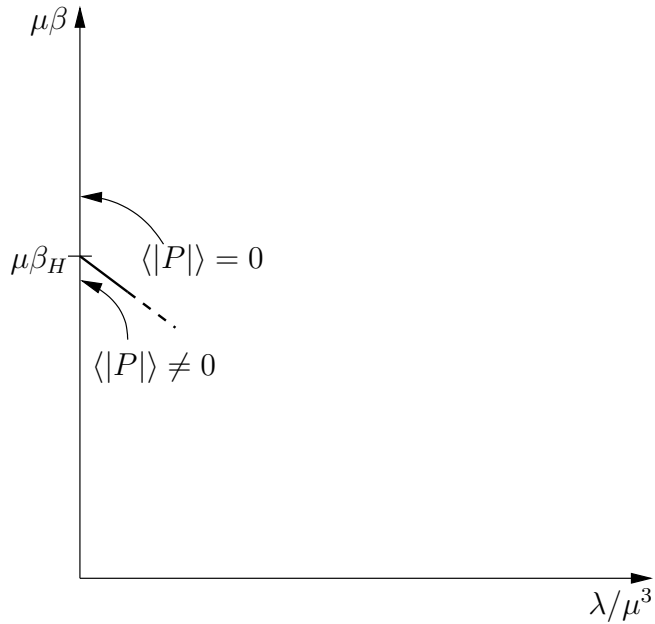}
  \caption{Schematic picture of the phase diagram of the PWMM at weak coupling.
  For $\mu\beta$ above $\mu\beta_H$ (at low temperature) the theory is confined.
  As $\mu\beta$ is lowered through $\mu\beta_H$ the theory undergoes a
  Hagedorn/deconfinement transition. The transition temperature can be computed
  at weak coupling; the calculation in the quenched theory is reviewed in
  section \ref{sec:quwk}, and section \ref{sec:ftwk} presents the result in
  the full theory.}
  \label{fig:schfd}
  \end{center}
\end{figure}

The existence of results at weak coupling suggests where we should look for
the phase transition at strong coupling. A schematic picture of the phase
diagram at weak coupling is shown in figure \ref{fig:schfd}. We will see below
that the weak coupling analysis suggests that $\mu\beta_H$ decreases as the
coupling is turned on. As a result, for our simulations, we will pick some
fixed $\mu\beta<\mu\beta_H$ and look for the transition by picking various
values of the coupling constant. We will first examine what happens in the
quenched theory, and thereafter present results for the full theory.

\subsection{Quenched Theory}\label{sec:quenched}
Here we will first present results for the quenched theory at weak coupling,
which can be culled from \cite{fss,mark+sv}, and then the results of our
simulations.
\subsubsection{Weak Coupling}\label{sec:quwk}
By counting gauge invariant states, it was argued in \cite{fss} that
a matrix model with action
\begin{equation}\label{Stoy}
  S=\int dt\sum_{j=1}^d \frac12\Tr\left[\left(\frac{d}{dt} X_j+i[A,X_j]\right)^2
  -\omega_j^2 X_j^2\right] ,
\end{equation}
with bosonic matrices $X_j$ in the adjoint of $SU(N)$, and $A$ being an $SU(N)$
gauge field, has a partition function that, in the large $N$ limit, is
approximately
\begin{equation}\label{Zquwk}
  Z(\beta)\approx \frac{e}{1-\sum_{j=1}^d e^{-\beta\omega_j}} .
\end{equation}
This partition function clearly diverges when the denominator vanishes, and it
was argued in \cite{fss} that the temperature where this happens is the Hagedorn
temperature.
\begin{equation}
  \sum_{j=1}^d e^{-\beta_H\omega_j}=1 .
\end{equation}

If we consider the various fields in the PWMM and decompose them into
representations of $SU(2|4)$ which contains a bosonic $SU(2)\times SU(4)$,
there are scalars that transform as a singlet under $SU(4)$ and in the $3$ of
$SU(2)$ and scalars that transform as a singlet under $SU(2)$ and in the $6$ of
$SU(4)$. If we expand the quenched theory about the trivial vacuum state, we
then have three matrices with $\omega=\mu$ and six matrices with $\omega=\mu/2$.
Writing $x_H=e^{-\beta_H\mu/2}$ we find the Hagedorn temperature is given by
the solution of
\begin{equation}
  3x_H^2+6x_H=1 ,
\end{equation}
where the admissible solution satisfies $0\le x_H \le1$. This gives
$x_H=-1+\frac2{\sqrt{3}}$, so that
\begin{equation}\label{mubqu}
  \mu \beta_H=-2\log\left(-1+\frac2{\sqrt{3}}\right)
  \approx 3.732528074 \, .
\end{equation}

Since the only massless field in \eqref{Stoy} is the holonomy of
the gauge field around the thermal circle, it is possible to integrate out
the matrices to find an effective action. This effective action is just the
effective action for the Polyakov loop. At one-loop level it was shown in
\cite{fss} that there is a deconfinement transition, whose temperature
coincides with the Hagedorn temperature.

It is also possible to calculate the effective action of the Polyakov loop
to higher order in the coupling constant. This has been carried out in the
full theory to two-loop order in \cite{mark+sv}, and three-loop order
in \cite{hrsy}. The two-loop calculation is sufficient to determine the
correction to the critical temperature, whereas three-loops are needed to
determine the order of the phase transition.

The next order correction to the Hagedorn temperature comes from the planar
two-loop effective action for the Polyakov loop. In this case, using the
results of \cite{mark+sv}, the partition function gets corrected
to\footnote{There is also a correction of order $\tilde\lambda$ coming from
the non-planar diagrams, but it does not contribute to the correction to
the Hagedorn temperature, so we don't include it.}
\begin{equation}
  Z=\frac{e}{1-3y^4-6y^2-\tilde\lambda g(y)\log y} ,
\end{equation}
with
\begin{equation}
  g(y)=48y^2(y^6+3y^4+7y^2+13) ,
\end{equation}
where $y=e^{-\beta \mu/4}$, and we write a dimensionless 't Hooft coupling
$\tilde\lambda=\lambda/\mu^3$.
Let us define
\begin{equation}\label{zqu}
  z(y)=3y^4+6y^2 .
\end{equation}
The correction to the transition temperature is given by
\begin{equation}\label{yhqu}
  y_H=y_H^{(0)}(1- c\ln y_H^{(0)}\tilde\lambda) ,
\end{equation}
or, equivalently,
\begin{equation}\label{bhqu}
  \mu\beta_H = -4\log(y_H^{(0)})(1-c\tilde\lambda) ,
\end{equation}
for a constant $c$ that is determined by
\begin{equation}\label{cqu}
  c=\frac{g(y_H^{(0)})}{y_H^{(0)} z'(y_H^{(0)})} .
\end{equation}
From \eqref{bhqu} it can be seen that $c>0$ decreases the inverse temperature as
the coupling is turned on, whereas $c<0$ increases it. Evaluating this at
$y_H^{(0)}=\sqrt{x_H}$ gives
\begin{equation}\label{cvalqu}
  c\approx 49.04614622 \, ,
\end{equation}
which means that the inverse temperature at the transition decreases as the
coupling is turned on.

\subsubsection{Strong Coupling}\label{sec:qust}
Here we will study the quenched theory by expressing all length scales in terms
of $\mu$. The quenched theory is useful for exploring some of the key issues
of our lattice methods. The PWMM describes interesting physics in various
limits in which the rank of the gauge group, $N$, is taken to infinity. Since
the phase space of the model grows very quickly with $N$ it is necessary to
understand if interesting physics is within the range of $N$ that is
computationally accessible. Moreover, we are also limited by the lattice
approximation itself and we would like to determine how many lattice points
are required to obtain a reasonable approximation to the continuum.

\begin{figure}
  \input{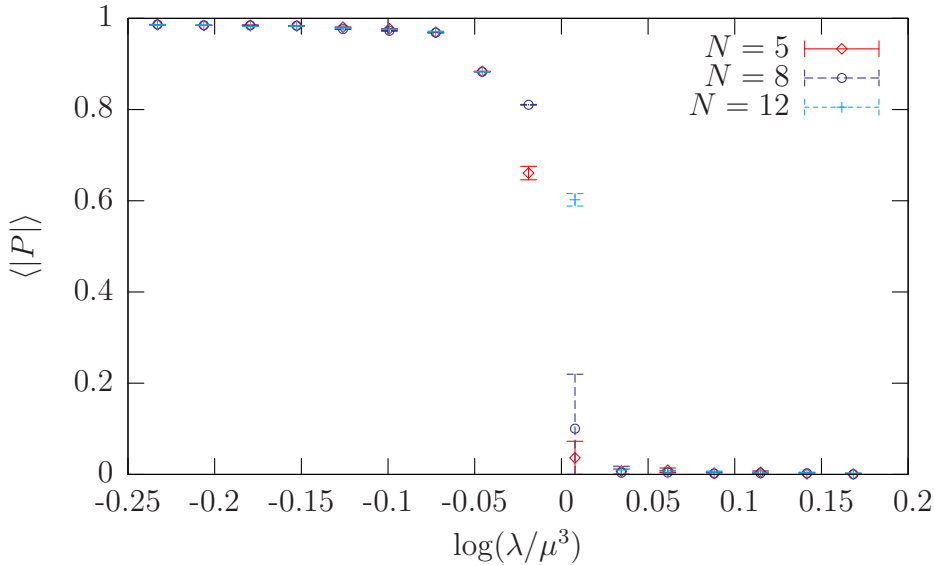}
  \caption{$\left<|\Tr e^{i\oint A}|\right>$ in the quenched theory for gauge
  group $SU(N)$ with various $N$, and $T=5$ lattice points. The results agree
  quite well for all choices of $N$ above the transition, and well below the
  transition. The greatest difference is near the transition, where it appears
  to be sharper in the case of larger $N$, as expected.}
  \label{fig:QGpolyN}
\end{figure}

Figure \ref{fig:QGpolyN} shows a plot of the Polyakov loop in the quenched
theory as a function of the lattice coupling measured in units of $\mu$, with
the inverse temperature $\beta\mu=1$, for various $N$ with five lattice points.
The plot shows clear evidence of a deconfinement transition. Moreover, it can be
seen from this plot that the critical value of the coupling does not seem to be
very dependent on $N$. This is very suggestive that the critical coupling in
the large $N$ theory can be reasonably approximated by simulations with modest
values of $N$. Note, however, the transition does appear to be much sharper for
larger $N$, as expected.

In figure \ref{fig:QGpolyT} we plot the Polyakov line in the quenched theory as
a function of the coupling at fixed $\beta\mu=1$ and $N=5$ for various lattice
sizes, $T$. The plot indicates that the critical value of the coupling does not
depend strongly on the lattice size, and that $T=5$ provides a reasonably good
approximation to the continuum.
\begin{figure}
  \input{QGpolyT.tex}
  \caption{$\left<|\Tr e^{i\oint A}|\right>$ in the quenched theory for
  various $T$, $N=5$.}
  \label{fig:QGpolyT}
\end{figure}

\begin{figure}
  \begin{center}
  \input{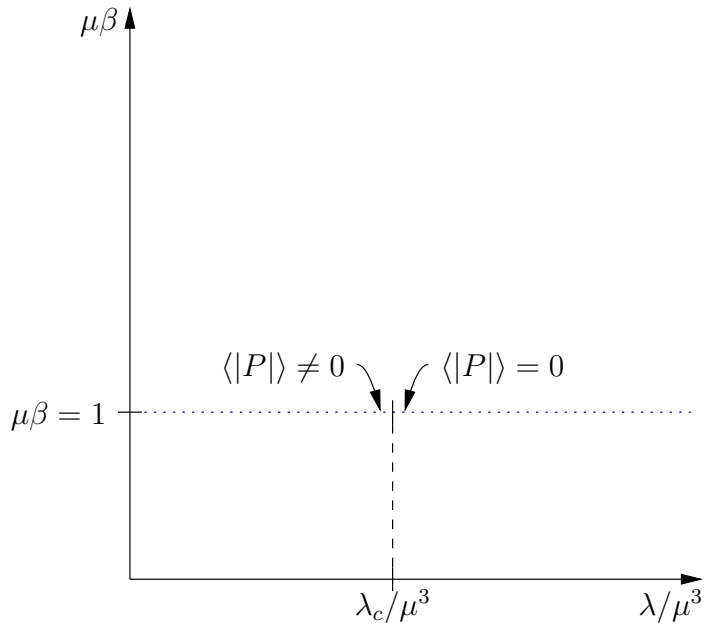}
  \caption{Schematic picture of the phase diagram from our lattice results.
  We study the theory along the blue dotted line, and identify a critical
  coupling, $\lambda_c$. At strong coupling (at this temperature) the theory
  is confined. It undergoes a deconfinement transition at $\lambda_c$. Compare
  this with the phase diagram at weak coupling in figure \ref{fig:schfd}.}
  \label{fig:phst}
  \end{center}
\end{figure}
Let us compare these results with those at weak coupling. If we extrapolate
the weak coupling results to $\mu\beta=1$, we can use \eqref{bhqu} to estimate
the critical value of the coupling. We find
$\log(\lambda/\mu^3)\approx-1.826043339$. Comparing this value to the plot
in figure \ref{fig:QGpolyN}, we see that the critical value of the coupling
differs considerably.

There are various possible reasons for this difference. One is that the weak
coupling computation was done in the 't Hooft limit expanding about the trivial
vacuum state. The results of our simulations, in contrast, should include the
effects of all of the various vacua, and are for finite $N$. Thinking about our
results from the weak coupling point of view, as we change $N$ our results
could be sensitive to both the number of oscillators above each vacuum state,
and the number of vacuum states, which change with $N$. We would expect that
both of these effects would show up in the $N$ dependence of our results, yet
figure \ref{fig:QGpolyN} shows the only prominent $N$ dependence is in the
sharpness of the transition. 

\subsection{Full Theory}\label{sec:full}
\subsubsection{Weak Coupling}\label{sec:ftwk}
The thermodynamics of multi-matrix models with fermions was also considered in
\cite{fss}. It was shown that for free gauged matrix quantum mechanics with
a collection of bosonic and fermionic matrices with frequencies $\omega_j$
the partition function in the large $N$ limit is approximately
\begin{equation}
  Z(\beta)\approx \frac{e}{1-\sum_{j=1}^d e^{-\beta\omega_j}} .
\end{equation}
I.e.\ \eqref{Zquwk} continues to hold, however the sum now runs over the
frequencies of both the bosonic and fermionic matrices. Again, the inverse
Hagedorn temperature is given by the place where the partition function
diverges. Moreover, as in the quenched case this can be shown to coincide with
the deconfinement transition temperature computed from the one loop effective
action for the Polyakov loop.

If we again consider classifying the fields by their transformations under
$SU(2|4)$, in addition to the bosonic scalars we had in the quenched case
we now also have fermions that transform in the fundamental representation of
both $SU(2)$ and $SU(4)$. If we expand about the trivial vacuum state, then,
we have, in addition to the three bosonic matrices with $\omega=\mu$ and six
with $\omega=\mu/2$, eight fermionic matrices with $\omega=3\mu/4$. Writing
$y=e^{-\beta\mu/4}$ we find the Hagedorn temperature is given by the solution of
\begin{equation}
  3y_H^4+8y_H^3+6y_H^2-1=(3y_H-1)(y_H+1)^3=0 ,
\end{equation}
where $y=y_H$ at the transition and the admissible solution satisfies
$0\le y_H \le1$. This gives $y_H=\frac13$, so that
\begin{equation}\label{mubft}
  \mu \beta_H=4\log3
  \approx 4.394449156 \, .
\end{equation}
Comparison with the result in the quenched theory \eqref{mubqu} gives a
difference of about $15\%$.

As in the quenched case, the next order correction to the Hagedorn temperature
comes from the planar two-loop effective action for the Polyakov loop. Here,
again using results of \cite{mark+sv}, the partition function gets corrected to
\begin{equation}
  Z=\frac{e}{1-z(y)-\tilde\lambda g(y)\log y} ,
\end{equation}
with
\begin{equation}\label{zgft}
  z(y)=3y^4+8y^3+6y^2
  \qquad
  g(y)=48y^2(1+y)^4(1+y^2) \, .
\end{equation}
As in the quenched case, the correction to the Hagedorn temperature, written
in the form \eqref{bhqu} gives \eqref{cqu} with
\begin{equation}\label{cvalft}
  c=\frac{205}{81} \, .
\end{equation}
Comparing this to the result in the quenched theory, \eqref{cvalqu}, shows
that the correction to the inverse temperature of the transition is stronger
by about an order of magnitude in the quenched case compared to the full theory.

\subsubsection{Strong Coupling}\label{sec:ftst}
We study the PWMM at strong coupling using lattice techniques, as in the
quenched case. However, the inclusion of the fermionic Pfaffian substantially
increases the computational expense. Fortunately, the results of the quenched
simulations suggest that five lattice points provide a reasonable approximation
to the continuum, at the temperature we are considering.

\begin{figure}
  \input{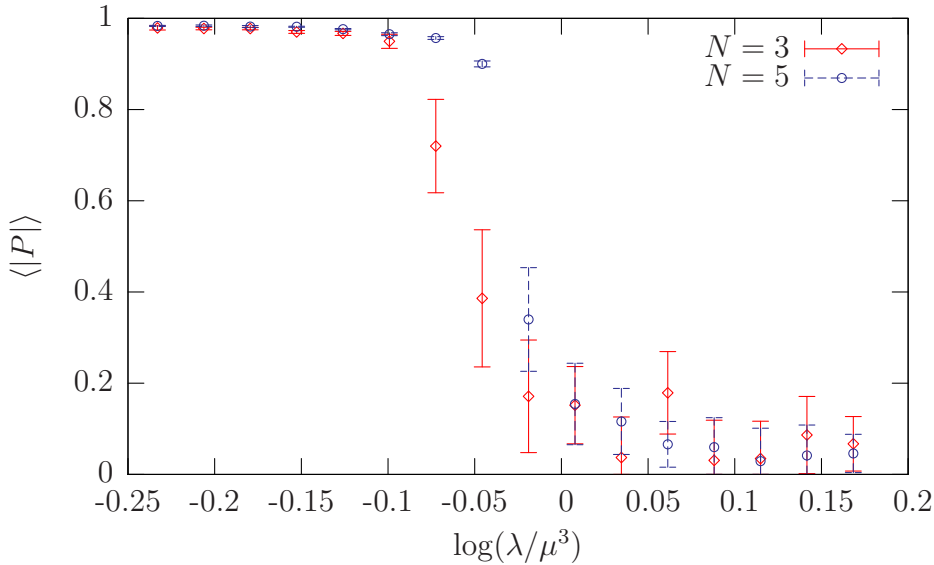}
  \caption{$\left<|\Tr e^{i\oint A}|\right>$ In the full theory with dynamical
  fermions. The gauge group is $SU(N)$ for $N=3,5$, and we are using five
  lattice points.}
  \label{fig:Tpoly}
\end{figure}
\begin{figure}
  \input{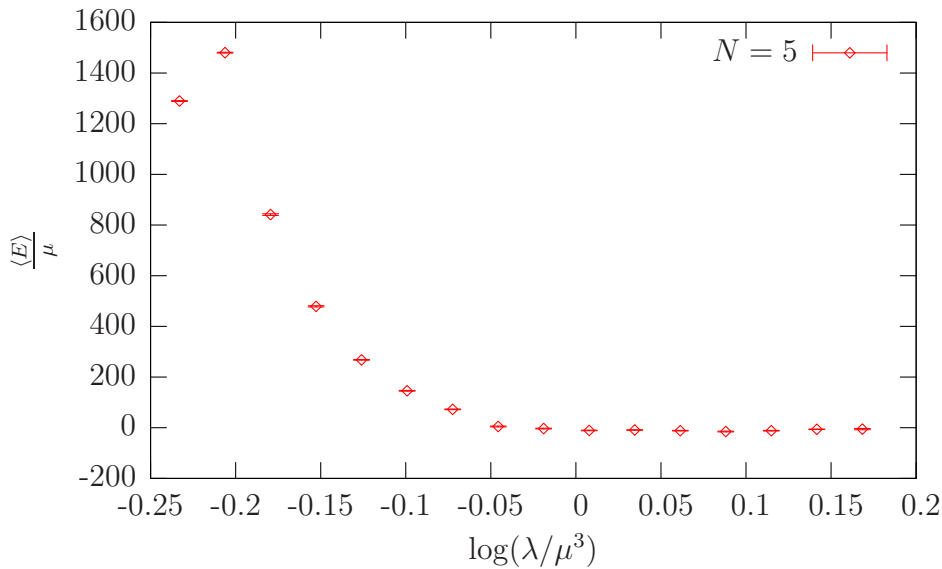}
  \caption{Energy $\frac{E}{\mu}$ vs $\log{\lambda/\mu^3}$ for $SU(5)$
  }
  \label{fig:Tact}
\end{figure}

Let us compare these results with those at weak coupling. If we extrapolate
the weak coupling results to $\mu\beta=1$, we can use \eqref{bhqu} to estimate
the critical value of the coupling. We find
$\log(\lambda/\mu^3)\approx-0.5154039782$. Comparing this value to the plot
in figure \ref{fig:Tpoly}, we see that the actual critical value of the coupling
differs considerably. 
Another interesting order parameter for the transition
is shown in figure~\ref{fig:Tact} which plots the 
expectation value of the energy in dimensionless
units as a function of the dimensionless 't Hooft coupling. This is
essentially given by the value of the bosonic action since the
contributions of the fermions can be computed exactly by a scaling
argument. Clearly, for strong coupling or small mass the vacuum
is supersymmetric $\bigl<\frac{E}{\mu}\bigr>=0$ while 
supersymmetry apparently spontaneously breaks
for small 't Hooft coupling.

If we compare the critical coupling of the full theory as plotted in figure
\ref{fig:Tpoly} with that of the quenched theory plotted in figure
\ref{fig:QGpolyN} we see that there is little difference. This might be
surprising, considering that extrapolation of the weak coupling results
suggested they should differ by more than an order of magnitude. We will comment
on this in the next section.

\begin{figure}
  \input{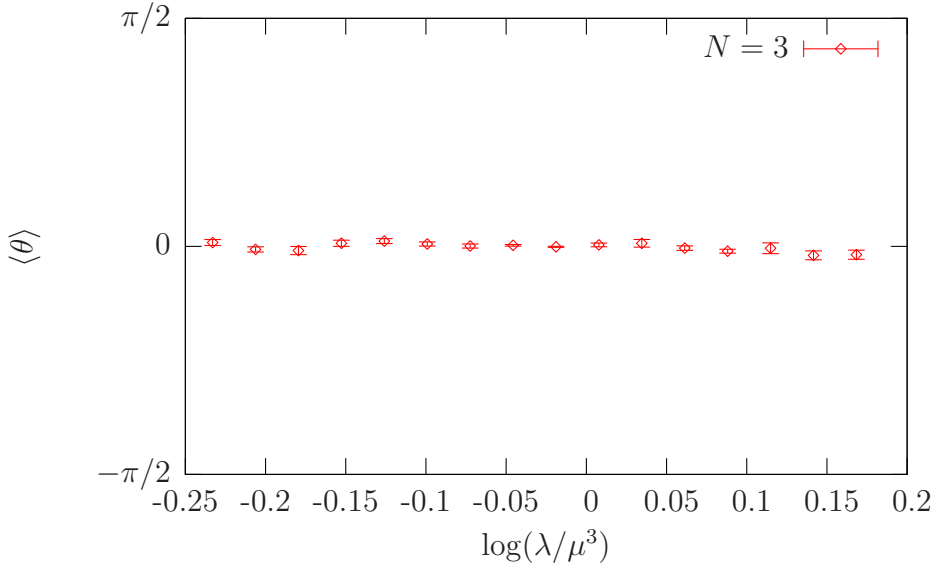}
  \caption{The phase of the Pfaffian for the full theory with gauge group
  $SU(3)$, fixed $\mu\beta=1$, and five lattice points. The phase is clearly
  small in the region of parameter space we are exploring.}
  \label{fig:TPfaff}
\end{figure}

A potential pitfall of our numerical simulations is the notorious sign problem.
In figure \ref{fig:TPfaff} we plot the Pfaffian phase for the theory with
gauge group $SU(3)$. The expectation value of the phase is very close to zero,
as expected, but more importantly, the uncertainty in the phase is very small.

\section{Discussion}\label{sec:disc}
\begin{figure}
  \begin{center}
  \input{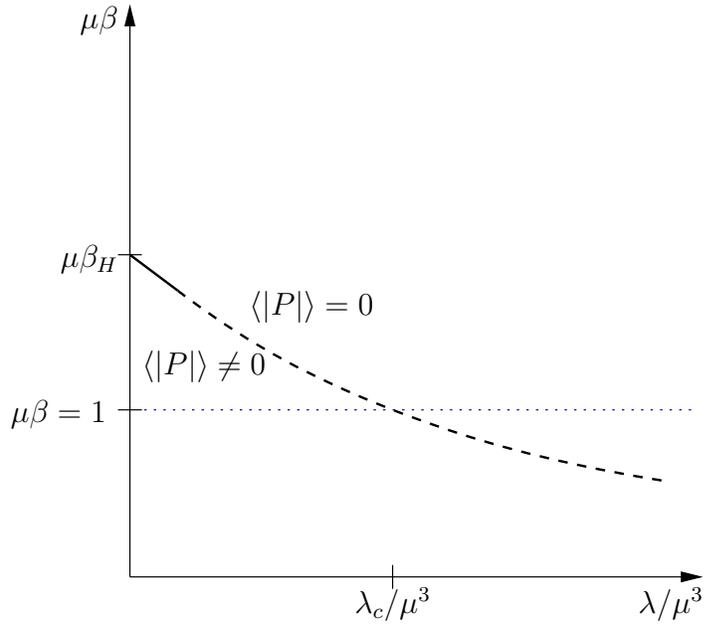}
  \caption{Hypothesized schematic picture of the phase diagram of the PWMM. Our
  simulation scans along the blue dotted line where we are able to identify a
  deconfinement transition. The heavy solid black line represents the known weak
  coupling result. The dashed black line represents a suggestion of what the
  phase diagram could look like interpolating between these points.}
  \label{fig:phsk}
  \end{center}
\end{figure}

We have presented results on lattice simulations of the PWMM. We argued that
the mass terms that deform the BFSS matrix model to the PWMM render it
particularly suitable for lattice study by lifting the moduli space and yielding
a discrete set of vacua. We have focussed on studying the Hagedorn/deconfinement
transition in the model. By simulating the model at fixed temperature over a
range of coupling, we have shown that the model exhibits a deconfinement
transition when the 't Hooft coupling is of order one. We found that critical
value of the coupling in the quenched case was of the same order as in the full
theory. At first glance this might be surprising because na\"ively extrapolating
the weak coupling result to the temperature at which we did our simulations
would suggest a difference of more than an order of magnitude. However, at
sufficiently high temperatures we expect the dynamics to be dominated by bosonic
zero modes on the thermal circle \cite{hte}, in which case the critical coupling
of the full theory should converge to that of the quenched theory. The fact that
our results seem to be indicate this convergence is setting in at $\mu\beta=1$
suggests that in future investigation of the phase diagram, it is sufficient to
study the quenched theory if we are interested in inverse temperatures that sit
below $\beta\mu=1$ in figure \ref{fig:phsk}.

In principle, the thermal ensemble should sample states near each of the
discrete vacua. In the large $N$ limit, since the number of vacua grows
exponentially with $N$, we expect that the moduli space problem encountered in
simulations of the BFSS model will reappear in the PWMM. The effect of the
additional vacua can be enhanced or suppressed depending on the physical
question of interest using umbrella sampling. Though we focussed on the
deconfinement transition, and the Polyakov loop is degenerate for non-trivial
vacua, we did not seem to need umbrella sampling to restrict to the trivial
vacuum. A full study of the behaviour of the scalar fields is interesting,
however, and we leave this for future work.

Though our simulation does not cover the entire $\beta-\lambda$
plane, it is tempting to speculate that the phase diagram is of the form in
figure \ref{fig:phsk}.
This seems most natural from the point of view of gauge/gravity duality. If
the transition at strong coupling was not connected to that at weak coupling,
for example, that would suggest the existence of some new gravity solution.

We have focussed on one particular aspect of the finite temperature strong
coupling dynamics of the PWMM, however, there are many other questions of
interest; we will mention two. In the limit of $\mu\to0$ at fixed $\beta$ and
$\lambda$, the PWMM becomes the BFSS matrix model. This model has a well-defined
dual black hole solution, and it would be interesting to explore this limit to
determine if the anticipated black hole thermodynamics is reproduced by the
strongly coupled matrix model. Also of interest is to simulate the theory in the
limit in which the PWMM becomes $\Ncal=4$ SYM on $R\times S^3$; results of
non-lattice techniques in this direction were presented in \cite{nishtalk}.

\section*{Acknowledgements}
We extend thanks to Mark Van Raamsdonk for helpful comments, and to David
Berenstein and Moshe Rozali for facilitating the discussion that led to this
work. GvA would like to thank the hospitality of Syracuse University while this
work was being initiated. This work is supported in part by the US Department of
Energy under grant DE-FG02-95ER40899. Simulations were done on the Legato
cluster at the University of Michigan, and using USQCD resources at
Fermilab.

\providecommand{\href}[2]{#2}
\raggedright


\begin{thebibliography}{100}

\bibitem{adscft1}
J.~M. Maldacena, {\it The large {N} limit of superconformal field theories and
  supergravity},  {\em Adv. Theor. Math. Phys.} {\bf 2} (1998) 231--252,
  [\href{http://xxx.lanl.gov/abs/hep-th/9711200}{{\tt hep-th/9711200}}].

\bibitem{adscft2}
S.~S. Gubser, I.~R. Klebanov, and A.~M. Polyakov, {\it Gauge theory correlators
  from non-critical string theory},  {\em Phys. Lett.} {\bf B428} (1998)
  105--114, [\href{http://xxx.lanl.gov/abs/hep-th/9802109}{{\tt
  hep-th/9802109}}].

\bibitem{adscft3}
E.~Witten, {\it Anti-de {S}itter space and holography},  {\em Adv. Theor. Math.
  Phys.} {\bf 2} (1998) 253--291,
  [\href{http://xxx.lanl.gov/abs/hep-th/9802150}{{\tt hep-th/9802150}}].

\bibitem{hnt}
M.~Hanada, J.~Nishimura, and S.~Takeuchi, {\it {Non-lattice simulation for
  supersymmetric gauge theories in one dimension}},  {\em Phys. Rev. Lett.}
  {\bf 99} (2007) 161602, [\href{http://xxx.lanl.gov/abs/0706.1647}{{\tt
  arXiv:0706.1647}}].

\bibitem{ahnt}
K.~N. Anagnostopoulos, M.~Hanada, J.~Nishimura, and S.~Takeuchi, {\it {Monte
  Carlo studies of supersymmetric matrix quantum mechanics with sixteen
  supercharges at finite temperature}},  {\em Phys. Rev. Lett.} {\bf 100}
  (2008) 021601, [\href{http://xxx.lanl.gov/abs/0707.4454}{{\tt
  arXiv:0707.4454}}].

\bibitem{cw2}
S.~Catterall and T.~Wiseman, {\it {Black hole thermodynamics from simulations
  of lattice Yang-Mills theory}},  {\em Phys. Rev.} {\bf D78} (2008) 041502,
  [\href{http://xxx.lanl.gov/abs/0803.4273}{{\tt arXiv:0803.4273}}].

\bibitem{cw3}
S.~Catterall and T.~Wiseman, {\it {Extracting black hole physics from the
  lattice}},  \href{http://xxx.lanl.gov/abs/0909.4947}{{\tt arXiv:0909.4947}}.

\bibitem{bmn}
D.~Berenstein, J.~M. Maldacena, and H.~Nastase, {\it Strings in flat space and
  pp waves from $\mathcal{N}=4$ super {Y}ang {M}ills},  {\em JHEP} {\bf 04}
  (2002) 013, [\href{http://xxx.lanl.gov/abs/hep-th/0202021}{{\tt
  hep-th/0202021}}].

\bibitem{lm}
H.~Lin and J.~M. Maldacena, {\it Fivebranes from gauge theory},  {\em Phys.
  Rev.} {\bf D74} (2006) 084014,
  [\href{http://xxx.lanl.gov/abs/hep-th/0509235}{{\tt hep-th/0509235}}].

\bibitem{inst}
G.~van Anders, {\it General {L}in-{M}aldacena solutions and {PWMM} instantons
  from supergravity},  {\em JHEP} {\bf 03} (2007) 028,
  [\href{http://xxx.lanl.gov/abs/hep-th/0701277}{{\tt hep-th/0701277}}].

\bibitem{lmsvv}
H.~Ling, A.~R. Mohazab, H.-H. Shieh, G.~van Anders, and M.~Van~Raamsdonk, {\it
  Little string theory from a double-scaled matrix model},  {\em JHEP} {\bf 10}
  (2006) 018, [\href{http://xxx.lanl.gov/abs/hep-th/0606014}{{\tt
  hep-th/0606014}}].

\bibitem{N4PWMM}
T.~Ishii, G.~Ishiki, S.~Shimasaki, and A.~Tsuchiya, {\it {$\mathcal{N}=4$ Super
  Yang-Mills from the Plane Wave Matrix Model}},  {\em Phys. Rev.} {\bf D78}
  (2008) 106001, [\href{http://xxx.lanl.gov/abs/0807.2352}{{\tt
  arXiv:0807.2352}}].

\bibitem{Phys_Rep_Exact_SUSY}
S.~Catterall, D.~B. Kaplan, and M.~Unsal, {\it {Exact lattice supersymmetry}},
  {\em Phys. Rept.} {\bf 484} (2009) 71--130,
  [\href{http://xxx.lanl.gov/abs/0903.4881}{{\tt arXiv:0903.4881}}].

\bibitem{bfss}
T.~Banks, W.~Fischler, S.~H. Shenker, and L.~Susskind, {\it M theory as a
  matrix model: A conjecture},  {\em Phys. Rev.} {\bf D55} (1997) 5112--5128,
  [\href{http://xxx.lanl.gov/abs/hep-th/9610043}{{\tt hep-th/9610043}}].

\bibitem{dsv1}
K.~Dasgupta, M.~M. Sheikh-Jabbari, and M.~Van~Raamsdonk, {\it Matrix
  perturbation theory for {M}-theory on a {PP}-wave},  {\em JHEP} {\bf 05}
  (2002) 056, [\href{http://xxx.lanl.gov/abs/hep-th/0205185}{{\tt
  hep-th/0205185}}].

\bibitem{dsv2}
K.~Dasgupta, M.~M. Sheikh-Jabbari, and M.~Van~Raamsdonk, {\it Protected
  multiplets of {M}-theory on a plane wave},  {\em JHEP} {\bf 09} (2002) 021,
  [\href{http://xxx.lanl.gov/abs/hep-th/0207050}{{\tt hep-th/0207050}}].

\bibitem{itt}
G.~Ishiki, Y.~Takayama, and A.~Tsuchiya, {\it $\mathcal{N}=4$ {SYM} on
  ${R}\times{S}^3$ and theories with 16 supercharges},  {\em JHEP} {\bf 10}
  (2006) 007, [\href{http://xxx.lanl.gov/abs/hep-th/0605163}{{\tt
  hep-th/0605163}}].

\bibitem{istt}
G.~Ishiki, S.~Shimasaki, Y.~Takayama, and A.~Tsuchiya, {\it Embedding of
  theories with ${SU}(2|4)$ symmetry into the plane wave matrix model},  {\em
  JHEP} {\bf 11} (2006) 089,
  [\href{http://xxx.lanl.gov/abs/hep-th/0610038}{{\tt hep-th/0610038}}].

\bibitem{N4PWMM2}
G.~Ishiki, S.-W. Kim, J.~Nishimura, and A.~Tsuchiya, {\it {Deconfinement phase
  transition in $\mathcal{N}=4$ super Yang-Mills theory on $R\times S^3$ from
  supersymmetric matrix quantum mechanics}},  {\em Phys. Rev. Lett.} {\bf 102}
  (2009) 111601, [\href{http://xxx.lanl.gov/abs/0810.2884}{{\tt
  arXiv:0810.2884}}].

\bibitem{N4PWMM3}
G.~Ishiki, S.-W. Kim, J.~Nishimura, and A.~Tsuchiya, {\it {Testing a novel
  large-N reduction for $\mathcal{N}=4$ super Yang-Mills theory on $R\times
  S^3$}},  {\em JHEP} {\bf 09} (2009) 029,
  [\href{http://xxx.lanl.gov/abs/0907.1488}{{\tt arXiv:0907.1488}}].

\bibitem{rhmc}
M.~A. Clark, A.~D. Kennedy, and Z.~Sroczynski, {\it Exact 2+1 flavour rhmc
  simulations},  {\em Nucl. Phys. Proc. Suppl.} {\bf 140} (2005) 835--837,
  [\href{http://xxx.lanl.gov/abs/hep-lat/0409133}{{\tt hep-lat/0409133}}].

\bibitem{jegerlehner}
B.~Jegerlehner, {\it {Krylov space solvers for shifted linear systems}},
  \href{http://xxx.lanl.gov/abs/hep-lat/9612014}{{\tt hep-lat/9612014}}.

\bibitem{metropolis}
N.~Metropolis, A.~W. Rosenbluth, M.~N. Rosenbluth, A.~H. Teller, and E.~Teller,
  {\it {Equation of state calculations by fast computing machines}},  {\em J.
  Chem. Phys.} {\bf 21} (1953) 1087--1092.

\bibitem{hmc}
S.~Duane, A.~D. Kennedy, B.~J. Pendleton, and D.~Roweth, {\it {Hybrid Monte
  Carlo}},  {\em Phys. Lett.} {\bf B195} (1987) 216--222.

\bibitem{hp}
S.~W. Hawking and D.~N. Page, {\it {Thermodynamics of Black Holes in anti-De
  Sitter Space}},  {\em Commun. Math. Phys.} {\bf 87} (1983) 577.

\bibitem{wittenconf}
E.~Witten, {\it Anti-de {S}itter space, thermal phase transition, and
  confinement in gauge theories},  {\em Adv. Theor. Math. Phys.} {\bf 2} (1998)
  505--532, [\href{http://xxx.lanl.gov/abs/hep-th/9803131}{{\tt
  hep-th/9803131}}].

\bibitem{ammpv}
O.~Aharony, J.~Marsano, S.~Minwalla, K.~Papadodimas, and M.~Van~Raamsdonk, {\it
  {The Hagedorn / deconfinement phase transition in weakly coupled large N
  gauge theories}},  {\em Adv. Theor. Math. Phys.} {\bf 8} (2004) 603--696,
  [\href{http://xxx.lanl.gov/abs/hep-th/0310285}{{\tt hep-th/0310285}}].

\bibitem{fss}
K.~Furuuchi, E.~Schreiber, and G.~W. Semenoff, {\it Five-brane thermodynamics
  from the matrix model},  \href{http://xxx.lanl.gov/abs/hep-th/0310286}{{\tt
  hep-th/0310286}}.

\bibitem{mark+sv}
M.~Spradlin, M.~Van~Raamsdonk, and A.~Volovich, {\it Two-loop partition
  function in the planar plane-wave matrix model},  {\em Phys. Lett.} {\bf
  B603} (2004) 239--248, [\href{http://xxx.lanl.gov/abs/hep-th/0409178}{{\tt
  hep-th/0409178}}].

\bibitem{hrsy}
S.~Hadizadeh, B.~Ramadanovic, G.~W. Semenoff, and D.~Young, {\it Free energy
  and phase transition of the matrix model on a plane-wave},  {\em Phys. Rev.}
  {\bf D71} (2005) 065016, [\href{http://xxx.lanl.gov/abs/hep-th/0409318}{{\tt
  hep-th/0409318}}].

\bibitem{hte}
N.~Kawahara, J.~Nishimura, and S.~Takeuchi, {\it {High temperature expansion in
  supersymmetric matrix quantum mechanics}},  {\em JHEP} {\bf 12} (2007) 103,
  [\href{http://xxx.lanl.gov/abs/0710.2188}{{\tt arXiv:0710.2188}}].

\bibitem{nishtalk}
J.~Nishimura, {\it {Non-lattice simulation of supersymmetric gauge theories as
  a probe to quantum black holes and strings}},  {\em PoS} {\bf LAT2009} (2009)
  016, [\href{http://xxx.lanl.gov/abs/0912.0327}{{\tt arXiv:0912.0327}}].

\end{thebibliography}
\end{document}